\documentstyle[12pt]{article}

\def\ba{\begin{array}{c}}
\def\ea{\end{array}}

\def\ben{$$}
\def\een{$$}
\def\be{\begin{equation}}
\def\ee{\end{equation}}

\begin{document}

\titlepage

 \begin{center}
 \vspace*{1cm}


{\Large \bf Perturbation method for non-square Hamiltonians and
its application to polynomial oscillators }

\vspace{15mm}

Miloslav Znojil

\vspace{3mm}

\'{U}stav jadern\'e fyziky AV \v{C}R, 250 68 \v{R}e\v{z},
Czech Republic\\

\vspace{3mm}

e-mail: znojil@ujf.cas.cz

 \vspace{25mm}

\end{center}

 \vspace{5mm}

\section*{Abstract}

A remarkable extension of Rayleigh-Schr\"{o}dinger perturbation
method is found and described. Its $(N+q) \times
(N+1)-$dimensional Hamiltonians are assumed emerging during
quasi-exact constructions of bound states.  At all $q > 1$, the
role of the traditional single eigenvalue is taken over by an
energy/coupling $q-$plet. In a way circumventing both the
non-linearity and non-Hermiticity difficulties, the corrections
are defined by compact, $q-$dimensional matrix-inversion formulae.

 \vspace{25mm}

PACS 03.65.Ge

\newpage

\section{Introduction}

The existence and efficiency of several sufficiently simple
perturbation techniques represents one of the not entirely
negligible reasons why the abstract formalism of Quantum Theory
finds so numerous and impressive practical applications. Usually,
one is given a ``realistic" (i.e., quite often, fairly
complicated) Schr\"{o}dinger equation
 \ben
 (H - E)\,|\psi\rangle = 0
 \een
and succeeds in its approximative replacement by a significantly
simpler equation
 \ben
 \left (H^{(0)} - E^{(0)}\right )\,|\psi^{(0)}\rangle = 0
 \een
where
 \begin{itemize}
 \item [(1)]
 the difference between the ``correct" and ``approximate"
measurable quantities (here, between energies $E_n$ and
$E^{(0)}_n$ at a few level-indices $n$) proves small,
 \item [(2)]
 the latter, ``zero-order" problem is comparatively easy to solve, and
 \item [(3)]
a suitable (e.g., Rayleigh-Schr\"{o}dinger \cite{Messiah}) version
of the so called perturbation theory is found as a source of a
recipe for a systematic improvement of the approximation.

 \end{itemize}

 \noindent
In a broader perspective, one may keep the energy $E$ fixed and
switch to a modified Schr\"{o}dinger operator, $H - E = H'(E) - g
V = H'' - S(E,g)= \ldots$. An appropriately modified perturbation
theory might then serve as a tool for the determination of the
``fine-tuned" couplings $g$ \cite{sturm}  or, whenever necessary,
of the whole multiplets of the couplings $g_1, \,g_2, \ldots,
g_{q-1}$ which may even re-incorporate the energy itself by
putting, formally, $E \equiv -g_0$. In the latter setting the
``original" or ``realistic" problem quite often acquires the
``$q-$plet-eigenvalue" generalized form
 \be
 \left [
H - S({g_0},{g_1}, \ldots, g_{q-1}) \right ]\,|g_0,g_1, \ldots,
g_{q-1}\rangle = 0\,
 \label{Magyaridva}
 \ee
which will be considered and studied in what follows.

\section{Polynomial-oscillator sample of equation (\ref{Magyaridva})}

Differential Schr\"{o}dinger equation
 \be
 -\psi_{}''(x)+ V_{}(x)\,\psi_{}(x)=
  E\,\psi_{}(x)\,
  \label{SE}
 \ee
for a particle moving in a one-dimensional polynomial well
 \be
 V_{}(x)= g_1x^2+g_2x^4+\ldots+g_{2q+1}x^{4q+2},
  \ \ \ \ \ g_{2q+1} > 0
 \label{pot}
 \ee
may be used as an illustrative example. Its solutions prove often
tractable via the infinite Taylor series
 \be
\psi_{}(x)= \exp [- P(x)]
 \times\,\sum_{n=0}^{\infty} h_nx^{2n+p}\,
 \label{QEbar}
 \ee
where the integer $p = 0$ or $1$ characterizes the parity. It may
be re-interpreted as an angular momentum $\ell = p-1 = 0, 1,
\ldots$ in three dimensions, etc.

In place of the general exponent $P(x)$ we shall only use here a
polynomial
 \be
  P(x)=\sum_{k=0}^{q}\ \frac{x^{2k+2}}{2k+2}\ f_k\,.
  \label{exponent}
 \ee
The asymptotically optimal choice of its coefficients $f_k$ is
uniquely determined by the WKB condition
 \be
 V_{}(x)= [P'(x)]^2+ {\cal O}\left (x^{2q}\right )\,
 \label{potb}
 \ee
or, if you wish,
 \be
 g_{2q+1}=f_q^2, \ \ \  g_{2q}=2f_qf_{q-1}, \ \ldots,\ \ \
 g_{q+1}=\sum_{j=0}^qf_{q-j}f_j\,
 \label{asym}
 \ee
which fixes all the $q+1$ coefficients $f_k$ as functions of the
first $q+1$ dominant couplings $g_{2q+1}, \,g_{2q}, \,
\ldots\,,\,g_{q+1}$ in potential (\ref{pot}). Thus, we have
$f_0=\sqrt{g_1}$ for harmonic oscillator ($q=0$) or
$f_1=\sqrt{g_3}$ and $f_0=g_2/(2f_1)$ for the sextic anharmonic
oscillator ($q=1$) etc.

The insertion of (\ref{QEbar}) transforms our differential
Schr\"{o}dinger eq. (\ref{SE}) + (\ref{pot}) into the
infinite-dimensional matrix problem (\ref{Magyaridva}) where the
non-square matrix of the system possesses the $(q+2)-$diagonal
banded form,
 \be
  {{H}}-S({g_0},{g_1}, \ldots,
 g_{q-1})=
 \left(
  \begin{array}{cccccccc}
 B_{0} & C_0&  & & &&& \\
 A_1^{(1)}&B_{1} & C_1&  &  &&& \\
 \vdots&\ddots&\ddots&\ddots&&&&\\
 A_q^{(q)}& \cdots& A_q^{(1)} &
 B_{q} & C_q&    && \\
 &\ddots&&\ddots&\ddots&\ddots&&\\
 &&A_{m}^{(q)}& \cdots&A_{m}^{(1)} &
 B_{m} & C_{m}& \\
 &&&\ddots&&\ddots&\ddots&\ddots
 \end{array}
 \right )\,.
  \label{Magy}
 \ee
The $g_k-$dependence of the matrix elements is transparent,
 \begin{eqnarray}
   C_n = (2n+2)\,(2n+2{p}+1),\ \ \ \ \ \
  B_n = -g_0-{f}_0\,(4n+2{p} +{1}) \equiv A_n^{(0)}, \nonumber
\\
 A_n^{(1)} = -{f}_1\,(4n+2{p}-1) + {f}_0^2 -g_1, \ \ \ \ \ \ \
 A_n^{(2)} = -{f}_2\,(4n+2{p}-3) + 2 {f}_0{f}_1 -g_2,  \nonumber \\
 \ldots,\ \ \ \ \ \ \ \ \ \ \ \ \ \ \ \ \ \ \ \ \ \ \
 \ \ \ \ \ \ \ \ \ \ \ \ \ \ \ \ \ \ \ \ \ \ \  \label{elem2}\\
 A_n^{(q)} = -{f}_{q}\,(4n+2{p}+1-2q) +
 \left ({f}_0{f}_{q-1}+ {f}_1{f}_{q-2}+ \ldots
  +{f}_{q-1}{f}_0
 \right ) -g_{q}, \nonumber
 \\
\ \ \ \ \ \ \ \ \ \ \ \ \ \ \ \ \
  \ \ \ \ n = 0, 1, \ldots \, \nonumber
  \end{eqnarray}
(remember that $-g_0 \equiv E$ is the energy). We have
 \be
 S({g_0},{g_1}, \ldots, g_{q-1})=\sum_{\xi=1}^{q}\,g_{\xi-1}
 {\cal J}_\xi\,
 \ee
where the generalized vectorial eigenvalue $\vec{g}=\left \{
{g_0},{g_1}, \ldots, g_{q-1}\right \}$ is attached to a $q-$plet
of auxiliary one-diagonal unit-like infinite-dimensional matrices
${\cal J}_\xi$.

\subsection{Illustration: $q=0$ (harmonic oscillator)}

At $q=0$ we may re-scale the coordinate $x$ and set $V(x) = x^2$
and $g_1=f_0=1$ giving $\psi_{}(x)= \exp (- x^2/2) \times\,\ldots$
etc. From the formal secular equation
 \ben
  \det \left[{H}-S({g_0})\right ]=\det
 \left(
  \begin{array}{cccccccc}
 B_{0} & C_0&  & & &&& \\
 0&B_{1} & C_1&  &  &&& \\
 &\ddots&\ddots&\ddots&&&&
 \ea
 \right )=0, \ \ \ \ \ \ \ q = 0
 \een
we deduce that the harmonic-oscillator spectrum is reproduced once
we let the secular determinant vanish by setting one of its
diagonal elements $B_m$ equal to zero. Thus, the $q=0$ example may
be skipped as trivial in what follows.

\section{Magyari's finite-dimensional equations (\ref{Magyaridva})}

In a pioneering letter \cite{Magyari}, E. Magyari discovered the
existence of the so called quasi-exact \cite{Ushveridze} bound
states in some special cases of the general polynomial potential
wells (\ref{pot}). In our present language, his claim is that the
infinite-dimensional matrix-equation representation
(\ref{Magyaridva}) + (\ref{Magy}) of the polynomial oscillators
admits a rigorous and exact finite-dimensional truncation. We only
have to select the very specific truncation-compatible
``intermediate" coupling
 \be
 g_{q}  = -{f}_q\,(4N+2q+2p+1) +
 \left ({f}_0{f}_{q-1}+ {f}_1{f}_{q-2}+ \ldots
  +{f}_{q-1}{f}_0
 \right )
 \label{redefuj}
 \ee
in order to arrive at the finite-dimensional matrices in eq.
(\ref{Magyaridva}),
 \be
  {{H}}-S=
 \left(
  \begin{array}{cccccccc}
 B_{0} & C_0&  & & &&& \\
 A_1^{(1)}&B_{1} & C_1&  &  &&& \\
 \vdots&\ddots&\ddots&\ddots&&&&\\
 A_q^{(q)}& \cdots& A_q^{(1)} &
 B_{q} & C_q&    && \\
 &\ddots&&\ddots&\ddots&\ddots&&\\
 &&A_{N-2}^{(q)}& \cdots&A_{N-2}^{(1)} &
 B_{N-2} & C_{N-2}& \\
 &&&A_{N-1}^{(q)}&\cdots& \ddots &
 B_{N-1} & C_{N-1} \\
 &&&&\ddots&\ddots&\vdots&B_N\\
 &&&&&A_{N+q-2}^{(q)}&A_{N+q-2}^{(q-1)}&\vdots\\
 &&&&&&A_{N+q-1}^{(q)}&A_{N+q-1}^{(q-1)}
 \end{array}
 \right ).
 \label{hatted}
 \ee
At every choice of the integer $N = 0, 1, 2, \ldots$ it makes
sense to contemplate the next, ${\cal O}\left (x^{2q} \right )$
order of precision in our WKB formula (\ref{potb}). This leads to
eq. (\ref{redefuj}) and connects the value of the subdominant
coupling $g_q$ with the $x^{2N+p}-$dominated truncated wave
functions~(\ref{QEbar}).

We get $A_n^{(q)} = 4\,{f}_q\,(N+q-n)$ and reduce the original
infinite-dimensional linear algebraic system (\ref{Magyaridva}) to
its finite-dimensional subset which consists of the mere $N+q$
rows. As long as we assumed that $h_{N+1} = h_{N+2} = \ldots = 0$,
only the $N+1$ Taylor coefficients $h_0, h_1, \ldots, h_N$ remain
unknown and must be determined by eq. (\ref{Magyaridva}). In this
sense the latter equation is overcomplete and, in effect,
non-linear. The operator $H - S$ becomes represented by a
non-square, $(N+q)\times (N+1)-$dimensional matrix.

\subsection{Sextic quasi-exact oscillators
 ($q=1$ and all $N=0,1,\ldots$)}

At the first nontrivial $q=1$ we may re-scale the coordinate $x$
and pick up, say, the even parity $p=0$. This gives $V(x) =
x^6+2f_0x^4+(f_0^2-3-4N)x^2 $ and the bound states $\psi_{}(x)=
\exp (- x^4/4-f_0x^2/2) \times\,\ldots$. One arrives at the {\em
square-matrix} secular equation of the $N+1-$dimensional form
 \ben
  \det \left[{H}-S({g_0})\right ]=\det
\left(
  \begin{array}{ccccc}
 B_{0} & C_0&  & &  \\
 A_1^{(1)}&B_{1} & C_1&  &   \\
 &\ddots&\ddots&\ddots&\\
 && \ddots & B_{N-1} & C_{N-1} \\
 &&&A_{N}^{(1)}&B_N
 \end{array}
 \right )
 =0, \ \ \ \ \ \ \ q = 1\,.
 \een
The standard square-matrix techniques may be employed so that also
all the quasi-exact constructions of the sextic oscillators with
$q=1$ may be skipped here as a mere routine exercise.

\subsection{Explicit solution at $N=0$ and all $q$}

Once we contemplate the higher degrees $q=2, 3, \ldots$ in our
quasi-exact polynomial-oscillator Schr\"{o}dinger-Magyari
non-square-matrix bound-state problem (\ref{Magyaridva}) +
(\ref{hatted}), we must admit that their solution represents a
challenge. The only manifestly non-numerical example occurs at
$N=0$ where the solution remains elementary,
 \begin{eqnarray}
  E= -g_0={f}_0\,(2{p} +{1}) , \ \ \ \ \ \
 g_1 =  {f}_0^2-{f}_1\,(2{p}+3)  , \ \ \ \ \
 \ldots,\ \ \ \ \  \nonumber \\
 g_{q-1} =\left ({f}_0{f}_{q-2}+ {f}_1{f}_{q-3}+ \ldots
  +{f}_{q-2}{f}_0
 \right ) -{f}_{q-1}\,(2{p}+2q-3) , \ \ \ \ N = 0. \nonumber
  \end{eqnarray}
The asymptotically dominant couplings ${g_{2q+1}},{g_{2q}},
\ldots, g_{q+1}$ in the potential (\ref{pot}) remain freely
variable while the $q+1$ elements of the remaining set
${g_0},{g_1}, \ldots, g_{q}$ become fixed as their explicit
quadratic functions. One may preserve such a pattern in all the $N
\geq 1$ cases.

\subsection{A toy decadic example: $q=2$ and $N=2$.  \label{toy}}

Before we get to the core of our present message and formulate the
general perturbation method of solving eq. (\ref{Magyaridva}) at
the arbitrarily large integers $q \geq 2$ and $N\geq 1$, let us
emphasize that the {\em idea} of the method may be most easily
illustrated by its application to the elementary decadic models
with $q=2$ and with the re-scaled $x$ such that $f_2= \sqrt{g_5} =
1$. The two free parameters $f_1 \ (\equiv g_4/2)$ and $f_0 \
[\equiv (g_3-f_1^2)/2]$ define $g_2 = 2f_0f_1- 2p - 13$ [cf. eq.
(\ref{redefuj})] and exponent $P(x)$ [cf. eq. (\ref{exponent})].
They also occur in the matrix elements of eq. (\ref{Magyaridva})
which reads
 \be
 \left (
 \begin{array}{ccc}
  -g_0-f_0-2pf_0&2+4p &0 \\
  f_0^2-g_1-3f_1-2pf_1& -g_0-5f_0-2pf_0 &12+8p \\
 8 &  f_0^2-g_1-7f_1-2pf_1& -g_0-9f_0-2pf_0 \\
 0 & 4& f_0^2-g_1-11f_1-2pf_1
 \ea
 \right )
\left (
 \begin{array}{c}
  h_0 \\
  h_1\\
  h_2
 \ea
 \right )=0
 \label{priklad}
 \ee
and which should determine the two unknown eigenvalues $g_0$
(representing the negative energy) and $g_1$ (= the non-variable
coupling at $x^2$) as well as the two coefficients $h_0$ and $h_1$
in the wave function (note that $h_2=1$ is a mere fixed
normalization constant).


\section{Perturbation forms of Magyari's equations (\ref{Magyaridva})}

Studies of analytic potentials found a mathematical encouragement
in the Regge's complex angular momentum method \cite{Newton}. In
this context, our parity parameter could be prolonged to the {\em
complex} angular momentum $\ell \equiv p-1 \in I\!\!\!C$. Rather
unexpectedly, the resulting extension of the range of $p$ will
enable us to simplify the bound-state problem (\ref{Magyaridva}) +
(\ref{hatted}) in the limit of the large $|p|\gg 1$ \cite{papers}.
Let us now outline the two alternative possibilities of such a
simplification in some detail.

\subsection{An elementary large$-\ell$ model}

In the toy example of paragraph \ref{toy} let us introduce the
quantity $\lambda =1/(2p)$ and assume that the numerical value of
this $\lambda$ may be chosen arbitrarily small. This implies that
the identical decomposition
 \be
 H=2p
 \left [ H^{(0)} + \lambda\,H^{(1)} \right ], \ \ \ \ \ q = N = 2
 \label{specia}
 \ee
of our Hamiltonian in eq.~(\ref{priklad}) with
 \ben
 H^{(0)}=
  \left (
 \begin{array}{ccc}
  -f_0&2&0 \\
  -f_1& -f_0 &4 \\
 0 &  -f_1& -f_0 \\
 0 & 0& -f_1
 \ea
 \right )  \ \ \ {\rm and}  \ \ \
  H^{(1)}=
  \left (
 \begin{array}{ccc}
  -f_0&2 &0 \\
  f_0^2-3f_1& -5f_0 &12 \\
 8 &  f_0^2-7f_1& -9f_0 \\
 0 & 4& f_0^2-11f_1
 \ea
 \right )
  \een
acquires the manifestly perturbative character. The dominant
$H^{(0)}$ is a very simple tridiagonal matrix while the more
complicated and four-diagonal matrix $H^{(1)}$ remains tractable
as a mere perturbation. Thus, the current conditions of the
applicability of perturbation algorithms are satisfied. We may
expect that at the larger integers $q$ and/or $N$, the same
pattern will enable us to postulate the formal split of any given
non-square Hamiltonian $H$ in some formal Taylor series
 \be
 H = H^{(0)} + \lambda\,H^{(1)}
 + \lambda^2\,H^{(2)}
 +\ldots
  \label{pertha}
 \ee
where the choice of the ``sufficiently small" parameter $\lambda$
is arbitrary. Even in the present letter this quantity may be
re-defined as $\lambda =1/(2p+const)$, in the spirit of the
popular ``shifted large$-\ell$" expansions \cite{Omar}.

\subsection{A more advanced large$-\ell$ arrangement of the same model
\label{subtleties}}

Let us abbreviate $-g_0-2pf_0=2pE$, $f_0^2-g_1-2pf_1=2pF$ and
re-scale eq.(\ref{priklad}),
 \be
 \left [
 {\cal T}\,
 \left (
 \begin{array}{ccc}
  2pE-f_0&2+4p &0 \\
  2pF-3f_1& 2pE-5f_0 &12+8p \\
 8 &  2pF-7f_1& 2pE-9f_0 \\
 0 & 4& 2pF-11f_1
 \ea
 \right )
 {\cal S}
 \right ]\,
 \left [
 {\cal S}^{-1}
\left (
 \begin{array}{c}
  h_0 \\
  h_1\\
  h_2
 \ea
 \right )\right ]=0
 \label{prikladny}
 \ee
with the ``small" $\sigma=p^{-1/3}$ in
 \ben
 {\cal T}=
 \left (
 \begin{array}{cccc}
 \sigma^{-1}/(4p)&&&\\
 & \sigma^{-2}/(4p)&&\\
 && \sigma^{-3}/(4p)&\\
 &&& \sigma^{-4}/(4p)
 \ea
 \right )\ \ \ {\rm and} \ \ \
 {\cal S}=
 \left (
 \begin{array}{ccc}
 1&&\\
 &\sigma&\\
 &&\sigma^2
 \ea
 \right ).
 \een
This leads to another re-formulation of our problem
(\ref{prikladny}),
 \be
 \left [H - S(s,t) \right ]\,|s,t\rangle
 =0, \ \ \ \ \ \
 |s,t\rangle =
 {\cal S}^{-1}\left (
 \begin{array}{c}
  h_0 \\
  h_1\\
  h_2
 \ea
 \right ), \ \ \ \ s = -\frac{E}{2\sigma}, \ \ \ \ t = -\frac{F}{2\sigma^2}
 \label{prikladnyje}
 \ee
and to another perturbation arrangement (\ref{pertha}) of the
Hamiltonian where the new unperturbed Hamiltonian
 \be
 H^{(0)}= \left (
 \begin{array}{ccc}
  0 &1 & 0 \\ 0 &0 & 2 \\ 2 &0 & 0 \\ 0 &1 & 0 \ea \right )
 \label{32}
 \ee
is accompanied by the three non-vanishing one-diagonal matrix
correction coefficients in eq. (\ref{pertha}) with $\lambda \equiv
\sigma = p^{-1/4} \ll 1$,
 \ben
 H^{(1)}=-\frac{f_1}{4}\,
 \left (
 \begin{array}{ccc}
 0&0&0\\
 3&0&0\\
 0&7&0\\
 0&0&11
 \ea
 \right ), \ \ \ \ \
 H^{(2)}=-\frac{f_0}{4}\,
 \left (
 \begin{array}{ccc}
 1&0&0\\
 0&5&0\\
 0&0&9\\
 0&0&0
 \ea
 \right ), \ \ \
 H^{(3)}=\frac{1}{2}\,
 \left (
 \begin{array}{ccc}
 0&1&0\\
 0&0&6\\
 0&0&0\\
 0&0&0
 \ea
 \right ).
 \een
The eigenvector $|s,t\rangle$ as well as the eigenvalue matrix $S$
in eq. (\ref{prikladnyje}) should be also represented by the
infinite Taylor-series ansatzs
 \be
 S=
  S^{(0)} + \lambda\,S^{(1)} + \lambda^2\,S^{(2)}
 + \ldots\,
 \label{redultopr}
 \ee
and
 \be
 |s,t\rangle=
  |s^{(0)},t^{(0)}\rangle + \lambda\,|s^{(1)},t^{(1)}\rangle
 + \ldots\,.
 \label{redultopr}
 \ee
The key specific merit of such a sophisticated innovation lies in
an unexpected and very significant simplification of the
zero-order (i.e., $|p| \to \infty$ limit of)
Schr\"{o}dinger-Magyari equations. For more details we may either
refer to our older paper \cite{Kratzer} or to the Appendix~A where
a few more technical details are summarized. These details are
instructive since our particular $q=N=2$ zero-order concrete model
(\ref{32}) admits a straightforward generalization to all $N$, at
$q \leq 5$ at least~\cite{papers}.

\section{The evaluation of perturbation corrections }


The generic infinite-dimensional matrix equation
(\ref{Magyaridva}) may be solved, say, by the technique of the so
called vectorial continued fractions \cite{vcf}. Perturbation
re-arrangements of such a treatment of our problem  exist but
still carry a more or less purely numerical character
\cite{fixed}. Here, an entirely different approach is being
employed.

For the sake of definiteness, we shall only pay attention to the
Magyari's finite-dimensional constructions at arbitrary truncation
integers $N < \infty$. The scope of the method might be broader in
principle but we decided to consider the Magyari's problem only.
Indeed, due to the practical difficulties with its solution it
hardly possesses any applications beyond the first few $q$s {\em
or} $N$s at present.

As we already indicated, our present approach to the Magyari's
problem (\ref{Magyaridva}) (we may call it  ``perturbed" problem)
will be perturbative. We shall postulate
 \be
 |g_{0},g_1, \ldots, g_{q-1}\rangle=
 |g_0^{(0)},g_1^{(0)}, \ldots, g_{q-1}^{(0)}\rangle
 +\lambda\,
 |g_0^{(1)},g_1^{(1)}, \ldots, g_{q-1}^{(1)}\rangle +\ldots\,
 \ee
and employ the parallel expansion (\ref{redultopr}) of the
$q-$plet of eigenvalues which contains $S=S({g_0},{g_1}, \ldots,
g_{q-1})$ and $S^{(k)}=S({g_0}^{(k)},{g_1}^{(k)}, \ldots,
g_{q-1}^{(k)})$ with $k=0, 1, \ldots$.


We may re-write our ``perturbed" Schr\"{o}dinger equation
(\ref{Magyaridva}) order-by-order in $\lambda$. One expects that
for the study of similar equations, all the solutions are at our
disposal not only for the overcomplete zero-order equation
 \be
H^{(0)} \,|g_0^{(0)},g_1^{(0)}, \ldots, g_{q-1}^{(0)}\rangle
=
S({g_0}^{(0)},{g_1}^{(0)}, \ldots, g_{q-1}^{(0)})
\,|g_0^{(0)},g_1^{(0)}, \ldots, g_{q-1}^{(0)}\rangle
\label{Magyarinula}
 \ee
but also for its under-complete ``left-action" partner
 \be
\langle _\xi \langle g_0^{(0)},g_1^{(0)}, \ldots,
g_{q-1}^{(0)}|\,H^{(0)}
=
\langle _\xi \langle g_0^{(0)},g_1^{(0)}, \ldots, g_{q-1}^{(0)}|\,
S({g_0}^{(0)},{g_1}^{(0)}, \ldots, g_{q-1}^{(0)})\,.
 \label{levice}
 \ee
On the level of the first-order corrections we complement the
zero-order homogeneous equation (\ref{Magyarinula}) by the {\em
non-homogeneous} linear algebraic problem using an abbreviation
$|\ 0\rangle=|g_0^{(0)},g_1^{(0)}, \ldots,  g_{q-1}^{(0)}\rangle$,
 \be
 \left [ H^{(0)}- S^{(0)}\right ] |g_0^{(1)},g_1^{(1)}, \ldots,
 g_{q-1}^{(1)}\rangle= \left [  S^{(1)}-H^{(1)}\right ]
 \,|\ 0\rangle
 \,.
  \label{Magyarijedna}
 \ee
Similarly we add its second-order descendant
 \be
 \left [ H^{(0)}- S^{(0)}\right ] |g_0^{(2)},g_1^{(2)}, \ldots,
 g_{q-1}^{(2)}\rangle= \left [  S^{(1)}-H^{(1)}\right ]\,
 |\ 1\rangle
  + \left [  S^{(2)}-H^{(2)}\right ]
 \,
 |\ 0\rangle
\label{Magyaridruha}
 \ee
(with abbreviated $|\ 1\rangle= |g_0^{(1)},g_1^{(1)}, \ldots,
g_{q-1}^{(1)}\rangle$) and so on. All the equations in this
hierarchy have the same recurrent structure,
 \be
 \left [ H^{(0)}- S^{(0)}\right ] |\ k\rangle=
 |\,{\rm known}^{(k-1)}\rangle +
   S^{(k)}
 \,
 |\ 0\rangle\,
\label{Magyarigeneric}
 \ee
with ``input" $ |\,{\rm known}^{(0)}\rangle =-H^{(1)} \,|\
0\rangle$ at $k=1$, etc.

This completes our description of the overall strategy and
algorithm. In dependence on the particular features of $H^{(0)}$
and of its eigenstates, a few more detailed technical aspects of
our innovated recipe may be added in a more explicit form which is
$H^{(0)}-$dependent of course.

For our concrete construction of section \ref{subtleties} (which
admits various generalizations \cite{papers}), the technical
details are particularly rich and relevant and they contribute to
a perceivable enhancement of the efficiency of the practical
calculations. All of them may be found summarized in Apendix~B
which offers a slightly more formal reading.

\section{Discussion and summary}

\subsection{The next-to-elementary quasi-exact bound states}

In various applications of Quantum Mechanics people sometimes
forget that the occurrence of the compact and elementary ``user
friendly" wave functions is not restricted to the mere linear
harmonic oscillator $V(x)=x^2$. Due to Magyari \cite{Magyari} we
know that the similar exceptional bound states, say, of the form
 \be
\psi_{(QE)}(x)= x^{\rm const}\exp (- {\rm polynomial}(x)) \times
{\rm polynomial}(x)
 \label{QEa}
 \ee
are much more generic and may be generated by virtually any
polynomial potential at certain particular subsets of couplings
and energies.

Unfortunately, all his formally flawless generalizations of the
$q=0$ harmonic oscillator to $q = 1, 2, \ldots$ found applications
solely at $q=1$ (cf., e.g., \cite{Singh}). Certain $q = 2$ models
attracted attention just at the very small $N$ \cite{litera}. In
all the other cases, the unfortunate coincidence of the
nonlinearity and non-Hermiticity of $H$ proved deterring. In this
sense, our present new approach might re-vitalize interest in the
undeniable phenomenological merits of the quasi-exact models with
the larger $q$s and $N$s.

In this direction we already re-analyzed a few particular
non-square Magyari's equations (\ref{Magyaridva}) at the freely
variable dimensions $N$ and discovered, purely empirically, that
some of these equations exhibit certain unexpected
``user-friendly" features up to $q=5$ at least \cite{papers}.
Moreover, it has been established by several groups of authors
that in a way paralleling the exceptional $q=1$ oscillators, some
of their ``first nontrivial" $q=2$ descendants {\em also} admit
the existence of quasi-exact multiplets which may be arbitrarily
large~\cite{BBjpa}. Efficiently, their non-Hermiticity may be
tamed by an appropriate modification of the ``physical" inner
product in Hilbert space~\cite{Geyer}.

All these observations might revitalize interest of people in the
Magyari's non-square matrix problem (\ref{Magyaridva}). In our
present paper we demonstrated that in spite of its non-linearity,
rather surprisingly, this anomalous form of the Schr\"{o}dinger
equation still admits a systematic perturbative solution. We
believe that the resulting facilitated perturbative tractability
might make Magyari's oscillators significantly more attractive in
applications.

\subsection{Innovations in perturbation theory}

Perturbation analysis of nonlinear systems usually requires a
narrow specification of their nonlinearity (cf., e.g., \cite{chin}
for illustration). Hence, it was a nice feeling to reveal that the
nonlinearity which characterizes the Magyari's equations
(\ref{Magyaridva}) still does not prevent us from employing the
ideas of the Rayleigh-Schr\"{o}dinger perturbation construction of
their solutions.

We should remind the readers that the use of the perturbation
series is one of the most natural strategies, able to provide
closed formulae as well as the upper and/or lower bounds of the
measurable quantities \cite{Hall}. Equally well it seems suitable
for the study of the real and complex eigenvalues \cite{compl}.
This puts our present perturbative approach to Magyari's problem
into a broader context and comparison.

First of all, the specific merit and recommendation of our present
approach for future calculations may be seen in the rapid growth
of the complexity of the various non-perturbative alternative
methods with the growth of $N$. Indeed, just the first few
smallest integers $N$ seem to have been considered in the
``minimally nontrivial" $q=2$ context up to now \cite{decadici}.

Secondly, our present method looks comparatively friendly also
with respect to the growth of the integer $q$ which measures the
flexibility of the shape of the potential (\ref{pot}). In a way
complementing and extending our thorough perturbation study at
$q=1$ \cite{drumi}, we showed here that an apparent
incompatibility of the perturbation-expansion strategy with the
Magyari's equations at the larger $q$ proves false. We imagined
that even some quite strongly non-square matrix forms of the
Magyari's Hamiltonian matrices with $q > 2$ remain tractable when
{\em both} their left and right action are taken into account {\em
simultaneously}.

In this sense, a certain methodical gap has been filled by our
present paper. Still, several open questions survive. For example,
the finite-dimensional character of our present Hamiltonian
matrices $H$ might help to suppress the weight of some problems
with convergence which often mar the use of the perturbation
techniques for many infinite-dimensional $H$s \cite{skala}. In
particular, one might recollect the encouraging rigorous
``acceleration of convergence" results as obtained at the
``trivial" $q=1$ in ref. \cite{drumi}.

\section*{Acknowledgement}

The work supported by the Institutional Research Plan AV0Z10480505
and by the GA AS CR grant Nr. A1048302.

\newpage

\section*{Appendix A: Zero-order example (\ref{32})}

For the time being let us ignore the notation subtleties and
changes of notation of section \ref{subtleties} and assume simply
that a given toy unperturbed non-square Hamiltonian (\ref{32}) may
be assigned the generalized eigenvalues
 \ben
 S^{(0)}({g_0^{(0)}},{g_1^{(0)}})=
\left (
\begin{array}{ccc}
{g_0}^{(0)} & &  \\{g_1}^{(0)} &{g_0}^{(0)} &  \\
 &{g_1}^{(0)} & {g_0}^{(0)} \\
 & & {g_1}^{(0)} \ea
\right )\ \equiv\ {g_0}^{(0)}{\cal J}_1 + {g_1}^{(0)}{\cal J}_2\,
 \een
and eigenvectors
 \ben
  |g_0^{(0)},g_1^{(0)}\rangle= \left ( \ba
  h_0^{(0)} \\ h_1^{(0)}  \\ h_2^{(0)}
 \ea \right ).
  \een
The exhaustive analysis of this problem would require a lot of
space. Fortunately, it has already been performed elsewhere
\cite{Kratzer}, with the most relevant result being that one just
gets not more than two real solutions
 \be
|1,1\rangle= \left (
\begin{array}{c}
1\\1\\1 \ea \right ), \ \ \ |-2,-2\rangle= \left (
\begin{array}{c}
1\\-2\\1 \ea \right ).
 \ee
They may be interpreted, as we already mentioned, as a doublet of
the decadic quasi-exact large$-\ell$ bound states. In some
applications, one also needs to find all the complex
eigenvalues~\cite{Tanguy} . Although they would still be tractable
by our present perturbation recipe, we shall ignore these
solutions here as unphysical.

\subsection*{A.1. Left zero-order eigenvectors}

The {\em left} eigenvectors of our perturbed as well as
unperturbed Hamiltonians $H$ may be found of interest as solutions
compatible with the conjugate equation
 \be
 \langle _\xi \langle g_0,g_1, \ldots, g_{q-1}|\,H
 =
 \langle _\xi \langle g_0,g_1, \ldots, g_{q-1}|\, S({g_0},{g_1},
 \ldots, g_{q-1})\,.
  \label{jered}
 \ee
In a double-bra Dirac-like notation, these row vectors $\langle
_\xi \langle g_0,g_1, \ldots, g_{q-1}|$ are numbered by an
additional subscript $ \xi = 1, 2, \ldots,\xi_{max}$. Such a
convention emphasizes the ambiguity of the solutions.

Our special illustration $H^{(0)}$ of eq. (\ref{32}) offers the
respective symmetric and antisymmetric left eigenvector
 \be
\langle _1 \langle 1,1|\equiv (1,1,1,1), \ \ \ \ \langle _2
\langle 1,1|\equiv  (3,-1,1,-3)\,
 \label{jedjed}
 \ee
at $g_0=g_1=1$, or
 \be
\langle _1 \langle -2,-2|\equiv  (2,-1,-1,2), \ \ \ \ \langle _2
\langle
 -2,-2|\equiv  (0,1,-1,0)
 \label{mindva}
 \ee
at $g_0=g_1=-2$. Arbitrary superpositions of these doublets may be
employed instead. This reflects the incomplete (or rather {\em
``under-complete"}) character of eq.~(\ref{jered}).

\subsection*{A.2. Reduced zero-order eigenvectors}

At any $q>1$, one might prefer the choice of the row eigenvectors
with as many zeros as possible. It is obvious that the linearly
independent $q-$plets of the vectors of this type may be written
as products
 \be
  \langle _\xi \langle g_0^{(0)},g_1^{(0)}, \ldots, g_{q-1}^{(0)}|
 =
 \langle_{j}\langle \varrho
 (
 g_0^{(0)},g_1^{(0)}, \ldots, g_{q-1}^{(0)}
 )
 |\times \Pi_{j}, \ \ \ \ \ j = 1, 2, \ldots, q
 \ee
with $(N+1)\times (N+q)-$dimensional ``reduction" matrices
$\Pi_{j}$ (= transpositions of some of the ``unit-like" matrices
${\cal J}_{\xi'}$ with $\xi \neq j \neq \xi'$ in general).

In our example at $q=N=2$, the sample of vectors in eq.
(\ref{jedjed}) may be replaced by the mutually conjugate
``reducible" eigenvectors $\langle _3 \langle 1,1|\equiv(3,1,2,0)$
and $\langle _4 \langle 1,1|\equiv(0,2,1,3)$. Their explicit
reductions
 \ben
 \langle _3 \langle 1,1| = (3,1,2)\times
 \left (
 \begin{array}{cccc}
 1&0&0&0\\
 0&1&0&0\\
 0&0&1&0
 \ea
  \right ), \ \ \ \
 \langle _4 \langle 1,1| =(2,1,3)\times
 \left (
 \begin{array}{cccc}
 0&1&0&0\\
 0&0&1&0\\
 0&0&0&1
 \ea
  \right )
 \een
may be abbreviated as $\langle _3 \langle 1,1|=\langle _1\langle
\varrho(1,1)|\,\Pi_1$ and $\langle _4 \langle 1,1|=\langle
_2\langle \varrho(1,1)|\,\Pi_2$, respectively, with $\Pi_j = {\cal
J}_{j}^T$.

\subsection*{A.3. Linearly independent sets}

The reduced, $(N+1)-$dimensional ``$\varrho-$vectors" $\langle
_j\langle \varrho|$ may be interpreted as the left eigenvectors of
the $(N+1)\times (N+1)-$dimensional {\em square} matrices $\left [
\Pi_{j} \left (H_{0}- S\right ) \right ] $. These reduced
eigenvectors must exist and be nontrivial since all their
``secular" determinants vanish. In such a context, one may
sometimes need to guarantee the linear independence of the row
vectors at the first $q$ subscripts $j=\xi=1,2,\ldots,q$ (say, by
an appropriate re-numbering of their ``overcomplete" available
family).

Sometimes, we may have to make the choice of $\Pi_j = {\cal
J}_{j'}^T$ with $j \neq j'$ for this purpose. For example, the
mutually conjugate superpositions $\langle _3 \langle -2,-2|\equiv
(1,0,-1,1)$ and $\langle _4 \langle -2,-2|\equiv (1,-1,0,1)$ of
the states (\ref{mindva}) lead to the respective factorizations
 \ben
 \langle _3 \langle -2,-2| =(1,-1,1)\cdot
 \left (
 \begin{array}{cccc}
 1&0&0&0\\
 0&0&1&0\\
 0&0&0&1
 \ea
  \right ),
 \langle _4 \langle -2,-2| = (1,-1,1)\cdot
 \left (
 \begin{array}{cccc}
 1&0&0&0\\
 0&1&0&0\\
 0&0&0&1
 \ea
  \right )
 \een
with the same form of the reduced array $\langle _1 \langle
\varrho(-2,-2)|$. Such a degeneracy reflects merely the presence
of the random zeros in the antisymmetric vector in (\ref{mindva}).
It may be avoided easily since the relation $\langle _2 \langle
-2,-2|=(0,1,-1)\,\Pi_1 \equiv (1,-1,0)|\,\Pi_2$ gives the other
two alternative options for the second and safely linearly
independent reduced vector $\langle _2 \langle \varrho(-2,-2)| $.

\section*{Appendix B. A few technical aspects of our innovated
perturbation recipe}


\subsection*{B.1. The explicit evaluation of the corrections to the
perturbed energies and couplings \label{energies}}

As long as our unperturbed Hamiltonian $H^{(0)}$ occurs also in
eq. (\ref{levice}) which may be re-written in the form
 \be
 \langle_{j}\langle \varrho
 (
 g_0^{(0)},g_1^{(0)}, \ldots, g_{q-1}^{(0)}
 )
 |\, \Pi_{j}
\,\left (H^{(0)}- S^{(0)} \right ) = 0\,, \ \ \ \  \ \ \ j = 1, 2,
\ldots, q\,,
 \ee
we may multiply  eq. (\ref{Magyarigeneric}) by the set of matrices
$\Pi_{j}$ and by the independent vectors $\ \langle_{j} \langle
\varrho \left ( g_0^{(0)},g_1^{(0)}, \ldots, g_{q-1}^{(0)} \right
)\, | \ \equiv \ \langle_{j}\langle \varrho|\ $ from the left.
This leads to the system of relations
 \be
 \langle_{j}\langle \varrho|\,
  \Pi_{j}
 \, S^{(k)}\,|\,0\rangle=-
 \langle_{j}\langle \varrho|\,
  \Pi_{j}
 \,
  |\,{\rm known}^{(k-1)}\rangle \,,
  \ \ \ \  \ \ \ j = 1, 2,
 \ldots, q\,,\,
  \label{rekl}
 \ee
i.e., to an elementary $q-$dimensional matrix-inversion definition
of the $k-$th-order energies/couplings $g_{\xi-1}^{(k)}$,
 \be
 \sum_{\xi=1}^{q}\,
 {\cal F}_{j,\xi}^{({\rm known})}\,g_{\xi-1}^{(k)}
 = c_j^{({\rm known})}\,, \ \ \ \  \ \ \ j = 1, 2,
\ldots, q\,.
  \label{reklas}
 \ee
Here, we merely introduced abbreviations $c_j^{({\rm known})}= -
\langle_{j}\langle \varrho|\, \Pi_{j} \, |\,{\rm
known}^{(k-1)}\rangle $ and $ {\cal F}_{j,\xi}^{({\rm known})} =
\langle_{j}\langle \varrho|\,  \left [\Pi_{j} \, {\cal
J}_\xi\right ]\,|\,0\rangle$ for scalar products of two known
vectors in $N+1$ dimensions. Their calculation may be facilitated
by the evaluation of all the elements
 \ben
 \left [G_{j,\xi} \right ]_{mn}= \sum_{k=1}^{N+q}\,\left [\Pi_{j}
  \right ]_{mk}
 \left [{\cal
J}_\xi \right ]_{kn}\,\ \ \ \ \ \ \ m,n=0,1,\ldots,N\,
 \een
in a preparatory step. One should remember that all the elements
of both the non-square and sparse matrix factors are here equal to
$0$ or $1$. The matrix elements of the product $\Pi_{j} \, {\cal
J}_\xi$ will often vanish, therefore. In the generic case with
$\Pi_{j} = {\cal J}_j^T$ we even get
 \be
  \left (\,|j-\xi|
  \geq N+1\,
  \right ) \ \
  \Longrightarrow \ \ {\cal J}_j^T\, {\cal J}_\xi = 0\,.
  \ee
This means that the sparse matrix ${\cal F}$ of the
coupling-defining system (\ref{reklas}) has a band-matrix
structure and its inversion is easier.

\subsection*{B.2. Perturbed eigenvectors \label{333}}

Whenever the solution of the zero-order eq. (\ref{Magyarinula}) is
unique, the ambiguity of the vector $|\ k\rangle$ specified by the
$k-$th eq. (\ref{Magyarigeneric}) lies in an uncontrolled
admixture of the $(N+1)-$dimensional zero-order vector $|\
0\rangle$. Once we construct any (normalized) basis in the
corresponding $(N+1)-$dimensional space,
 \ben
 \left \{ |\ \beta_m\rangle \right \}_{m=0}^{N}
 \een
and assume that the above vector coincides, say, with its zeroth
element, $|\ 0\rangle \equiv |\ \beta_0\rangle$, we may simply
follow the textbooks and define the ``right" projector
 \ben
 Q_R= \sum_{m=1}^{N}\,
  |\ \beta_m\rangle \,\langle  \beta_m\ |
 \een
which mediates the standard Rayleigh-Schr\"{o}dinger normalization
of $|\ k\rangle = Q_R|\ k\rangle$.

In parallel, we have to recollect that we already employed $q$
independent rows of the $k-$th eq. (\ref{Magyarigeneric}) for the
specification of $S^{(k)}$ via eq. (\ref{reklas}). More
specifically, we multiplied eq. (\ref{Magyarigeneric}) by $q$
vectors $ \langle_{j}\langle \varrho|\, \Pi_{j} $ from the left.
As long as these vectors were chosen as linearly independent, they
span a $q-$dimensional subspace in the ``larger",
$(N+q)-$dimensional vector space. This allows us to assume that
the orthogonal complement of this subspace is spanned by an
$N-$plet of some $(N+q)-$dimensional vectors $ |\
\alpha_n\rangle$. These vectors define our second, ``left"
projector
 \ben
 Q_L= \sum_{n=1}^{N}\,
  |\ \alpha_n\rangle \,\langle  \alpha_n\ |\,.
 \een
We are now ready to replace our non-homogeneous algebraic  eq.
(\ref{Magyarigeneric}) by its subsystem
 \be
  Q_L\,\left [ H^{(0)}- S^{(0)}\right ] Q_R\, |\ k\rangle=
  Q_L|\,{\rm known}^{(k-1)}\rangle +
    Q_L\,S^{(k)} \, |\ 0\rangle\,
\label{Magyarigenericright}
 \ee
where the second term on the right-hand side has been made
``known" in the previous subsection. By construction, the
left-hand-side $N$ by $N$ matrix is invertible and we have
 \be
  |\ k\rangle= Q_R\,\frac{1}{Q_L\,\left [ H^{(0)}- S^{(0)}\right ] Q_R}
  \,Q_L\,
  \left ( |\,{\rm known}^{(k-1)}\rangle +
    S^{(k)} \, |\ 0\rangle \right )\,
\label{Mafinright}
 \ee
which is our final and compact ``generalized
Rayleigh-Schr\"{o}dinger" explicit formula for the $k-$th
correction to the wave function.

\subsection*{B.3. Left perturbed eigenvectors}

We have seen that in the Magyari-equation context, the left
vectors only played a role in their {\em unperturbed}, zero-order
form so that the description of their perturbation construction
might simply be skipped as redundant. Nevertheless, for the sake
of completeness, let us add a few remarks also on the $q-$plet of
the row (left) eigenvectors of the non-square Hamiltonians and on
all the space spanned by their superpositions
 \be
 \langle\langle \psi|=\sum_{\xi=1}^{q}\,C_\xi\,\cdot\,
 \langle _\xi \langle g_0,g_1, \ldots, g_{q-1}|\,.
 \ee
By assumption, they all satisfy the perturbed conjugate eq.
(\ref{jered}) and we may expand both their separate components and
the coefficients in the power series in $\lambda$. The combination
of these series
 \be
 \langle _\xi \langle g_0,g_1, \ldots, g_{q-1}|
 =
 \langle _\xi \langle g_0^{(0)},g_1^{(0)}, \ldots, g_{q-1}^{(0)}|
 +\lambda\,
 \langle _\xi \langle g_0^{(1)},g_1^{(1)}, \ldots, g_{q-1}^{(1)}|
 +\ldots\,
 \ee
and $ C_\xi=\sum_{k}\,\lambda^k\,C_\xi^{(k)}$ gives
 \be
 \langle\langle \psi| = \sum_{k}
\,\lambda^k\,\langle\langle \psi^{(k)}|, \ \ \ \ \  \ \ \ \ \ \
  \langle\langle
 \psi^{(k)}|=\sum_{\xi=1}^q \,\sum_{n=0}^k\,C_\xi^{(k-n)}\cdot
 \langle _\xi \langle n\,|
 \label{rozvoj}
 \ee
where, in a naturally simplified notation, $ \langle _\xi \langle
n\,|\ \equiv \ \langle _\xi \langle g_0^{(n)},g_1^{(n)}, \ldots,
g_{q-1}^{(n)}|$.

Let us now enter the most characteristic postulate of {\em all}
the Rayleigh-Schr\"{o}dinger-type perturbation recipes which
removes the ambiguity of the eigenvectors by the consequent
requirement of a {\em complete absence} of any zero-order
component $ \langle _\xi \langle 0\,|$ in any correction
$\langle\langle \psi^{(k)}|$ with $k \geq 1$.

In our present case this option leads to the two independent
consequences. Firstly, in the light of eq. (\ref{rozvoj}) we have
to put $ C_\xi^{(1)}= C_\xi^{(2)}=\ldots = 0$ [i.e., $n=k$ in eq.
(\ref{rozvoj})] and get just $ C_\xi \equiv C_\xi^{(0)}$.
Secondly, {\em all} the vectors $ \langle _\xi \langle n\,|$ must
be constructed as perpendicular to {\em all} $ \langle _j \langle
0\,|$ at {\em any} $n \geq 1$, i.e., in our notation of section
\ref{333},
 \be
  \langle _\xi
 \langle n\,| = \langle _\xi \langle n\,|\,Q_L\,,
 \ \ \
 n = 1, 2, \ldots
  \,, \ \ \ \ \ \ \ \
 \xi = 1, 2, \ldots, q\,.
 \ee
With this normalization we may now insert eq. (\ref{rozvoj}) in
eq. (\ref{levice}) and arrive at an analogue of eq.
(\ref{Magyarigeneric}). Of course, we may skip the repetition of
the reconstruction of $S^{(k)}$ and jump immediately to the
following $k \geq 1$ analogue of eq. (\ref{Magyarigenericright}),
 \be
 \langle_\xi \langle
 k\,|\,
  Q_L\,\left [ H^{(0)}- S^{(0)}\right ] Q_R=
 \langle_\xi\langle\,{\rm known}^{(k-1)}\,|\,Q_R, \ \ \ \ \ \ \ \
 \xi = 1, 2, \ldots, q\,.
\label{Magyarigenericleft}
 \ee
With
 \be
 \langle_\xi\langle\,{\rm known}^{(0)}\,|=
 \langle_\xi \langle
 0\,|\,
 \left [  S^{(1)}-H^{(1)}\right ]
 \ee
at $k=1$ etc, the trivial $N$ by $N$ inversion gives the final
formula
 \be
 \langle_\xi \langle
 k\,|
 = \langle_\xi\langle\,{\rm known}\,|\cdot
 Q_R\,\frac{1}{Q_L\,\left [ H^{(0)}- S^{(0)}\right ] Q_R}
  \,Q_L\,, \ \ \ \ \ \ \ \
 \xi = 1, 2, \ldots, q\,.
\label{Mafinrightb}
 \ee
The detailed account of our Magyari-inspired recurrent recipe is
completed.

\end{document}